\documentclass[%
mathleft,%
final,%
]{an}
\usepackage{graphicx}
\usepackage{times}
\begin{document}

\Pagespan{1}{}
\Yearpublication{2013}%
\Yearsubmission{2012}%
\Month{1}%
\Volume{334}%
\Issue{1}%
\DOI{This.is/not.aDOI}%

\title{Fundamental properties of lower main-sequence stars}

\author{Guillermo Torres\thanks{Corresponding author. 
  \email{gtorres@cfa.harvard.edu}}
}
\titlerunning{Fundamental properties}
\authorrunning{G. Torres}
\institute{Harvard-Smithsonian Center for Astrophysics, 60 Garden St.,
Cambridge MA 02138, USA}

\received{XXXX}
\accepted{XXXX}
\publonline{XXXX}

\keywords{binaries: eclipsing, binaries:spectroscopic,
stars:evolution, stars:late-type, starspots.}

\abstract{The field of exoplanet research has revitalized interest in
M dwarfs, which have become favorite targets of Doppler and transit
surveys.  Accurate measurements of their basic properties such as
masses, radii, and effective temperatures have revealed significant
disagreements with predictions from stellar evolution theory in the
sense that stars are larger and cooler than expected.  These anomalies
are believed to be due to high levels of activity in these stars.  The
evidence for the radius discrepancies has grown over the years as more
and more determinations have become available; however, fewer of these
studies include accurate determinations of the temperatures. The
ubiquitous mass-radius diagrams featured in many new discovery papers
are becoming more confusing due to increased scatter, which may be due
in part to larger than realized systematic errors affecting many of
the published measurements.  A discussion of these and other issues is
given here from an observer's perspective, along with a summary of
theoretical efforts to explain the radius and temperature anomalies.}

\maketitle

\section{Introduction}
\label{sec:introduction}

A common justification for studying late-type stars, aside from being
interesting objects in themselves, is that they dominate the stellar
population in our Galaxy by number: roughly 75\% of the points of
light in the sky are M dwarfs.  In recent years they have also become
attractive targets for exoplanet searches (Nutzman et al.\
\cite{Nutzman:08}; Law et al.\ \cite{Law:11a}; Sip\H{o}cz et al.\
\cite{Sipocz:11}; Barnes et al.\ \cite{Barnes:12}; and others),
particularly when looking for small planets in the habitable zones of
their parent stars. This has provided extra motivation for studying
them.  Both the Doppler signals and the transit signals are larger and
more easily detectable for planets around M dwarfs, and the lower
stellar luminosities mean that orbits in the habitable zone are closer
in, making transits more likely and more frequent when they do occur.

An important by-product of recent transit surveys has been the
discovery of many eclipsing binaries with M-dwarf components (e.g.,
Law et al.\ \cite{Law:11b}; Coughlin et al.\ \cite{Coughlin:11};
Harrison et al.\ \cite{Harrison:12}; Birkby et al.\ \cite{Birkby:12}).
These kinds of systems have traditionally been the most favorable for
determining the basic properties of late-type stars, including their
mass, radius, temperature, and luminosity. Unfortunately, however,
most newly discovered systems tend to be faint, and the bottleneck for
accurate determinations continues to be the spectroscopy.

Masses and radii in double-lined eclipsing binaries can be obtained
free of assumptions, as their derivation depends only on Newtonian
physics and geometry. Many determinations have been made also in
single-lined eclipsing binaries, but these are less fundamental as
they require previous knowledge of the primary mass or radius, or the
assumption of synchronous rotation of the primary and spin-orbit
alignment in circular orbits. Long-baseline interferometry has enabled
the measurement of absolute radii for increasing numbers of single
late-type stars for which the parallax is known, but masses for these
stars cannot be obtained dynamically, so they are typically estimated
from a mass-luminosity relation. These assumptions limit the
usefulness of single-lined eclipsing binaries and single stars for
testing models of stellar evolution.

\section{Measuring fundamental properties}
\label{sec:measuring}

The techniques for measuring absolute masses and radii of stars in
eclipsing binaries are straightforward and sufficiently well known
that we will dispense with a description here. For details the reader
is referred to the reviews by Andersen (\cite{Andersen:91}), or
Torres, Andersen \& Gim\'enez (\cite{Torres:10}).  The application of
these methods still requires care, though, if precisions (and
accuracies) of 3\% or better are to be obtained, as are generally
necessary for meaningful comparisons with stellar evolution theory for
low-mass stars. Temperature determinations are less fundamental. In
eclipsing binaries they are typically derived either through color
indices and empirical calibrations (if the reddening is known), or
from the luminosities and radii if the distance is known.
Spectroscopic temperature determinations remain difficult due to the
complexities of modeling molecular features in the atmospheres of cool
stars, not to speak of the fact that the observed spectra in binaries
are typically double-lined. Metallicities have also been challenging
to determine spectroscopically for similar reasons, although progress
is being made in calibrating composition in terms of optical and
near-infrared indices (e.g., Woolf et al.\ \cite{Woolf:05},
\cite{Woolf:06}, \cite{Woolf:09}; Rojas-Ayala et al.\ \cite{Rojas:10},
\cite{Rojas:12}; Reyl\'e et al.\ \cite{Reyle:11}). Photometric
calibrations using various color indices are also available (Bonfils
et al.\ \cite{Bonfils:05}; Casagrande et al.\ \cite{Casagrande:08};
Johnson et al.\ \cite{Johnson:09}, \cite{Johnson:12}; Schlaufman et
al.\ \cite{Schlaufman:10}).

\section{Discrepancies with models}
\label{sec:discrepancies}

Some of the observed properties of low-mass stars are known to
disagree with predictions from standard stellar evolution theory. Most
notably, the measured radii for M stars in binaries are larger than
indicated by models, typically by 5--10\% for the best measured
systems, or sometimes more depending on the model. Early indications
of this problem for M dwarfs were reported by Hoxie (\cite{Hoxie:70},
\cite{Hoxie:73}) and Lacy (\cite{Lacy:77}), and were subsequently
supported by additional accurate measurements by Popper
(\cite{Popper:97}), Clausen (\cite{Clausen:99}), and many others. More
recent determinations have removed any observational doubt that the
models do not fit the observations as well as they should. While this
radius problem has received the most attention, the effective
temperatures of low-mass stars are also in disagreement with theory,
although far fewer studies have shown this because temperatures are
more difficult to determine. Real stars tend to be too cool compared
to predictions, by about half as much as the radii, in relative terms.

For decades there were only two low-mass eclipsing binaries with
properties measured accurately enough to notice the problem: the
classical systems CM Dra ($\sim$0.22\,$M_{\odot}$; see Morales et al.\
\cite{Morales:09}) and YY Gem ($\sim$0.60\,$M_{\odot}$; Torres \&
Ribas \cite{Torres:02}).  Then came CU Cnc ($\sim$0.4\,$M_{\odot}$;
Ribas \cite{Ribas:03}) and GU Boo ($\sim$0.6\,$M_{\odot}$;
L\'opez-Morales \& Ribas \cite{Lopez-Morales:05}), and although many
others have been published since then, relatively few of these studies
can claim realistic uncertainties in the masses and radii under 3\%,
and show convincingly that systematic errors are under control.

In recent years it has become clear that these discrepancies with the
models are not confined to the M dwarfs (Torres et al.\
\cite{Torres:06}; Clausen et al.\ \cite{Clausen:09}); some stars as
massive as the Sun (or more generally, stars with convective
envelopes) also seem to be ``inflated'' and too cool when compared
with models.

\begin{figure}
\includegraphics[angle=0,width=\linewidth]{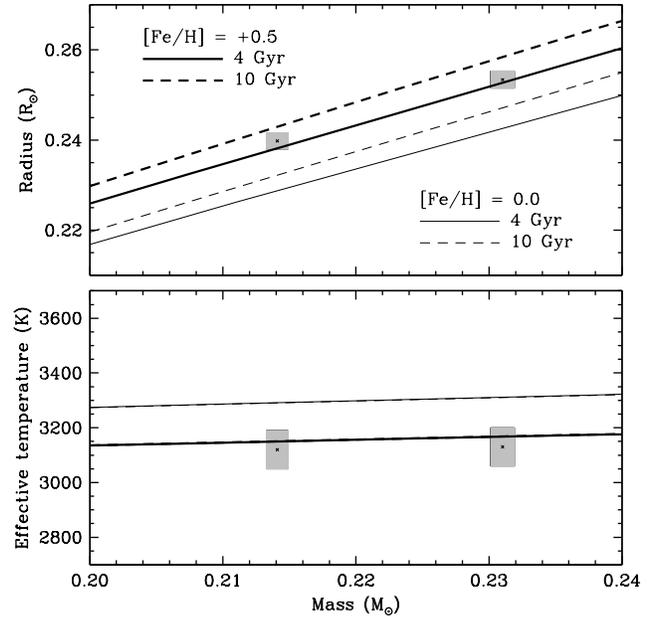}
\caption{Measured masses, radii, and effective temperatures for CM Dra
compared with isochrones by Dotter et al.\ (\cite{Dotter:08}). In the
lower panel the 4 Gyr and 10 Gyr isochrones at each metallicity are
indistinguishable.}
\label{fig:cmdra}
\end{figure}

The clearest evidence of ``radius inflation'' and ``temperature
suppression'' is seen by examining the best-studied individual
systems.  CM Dra is the poster child for these anomalies (see also
discussions by Feiden et al.\ \cite{Feiden:11}, Spada \& Demarque
\cite{Spada:12}, and MacDonald \& Mullan \cite{MacDonald:12}). It is
presumed to be a Population II binary based on its extreme kinematics,
although its precise age is not known. The metallicity has also been
controversial, with recent determinations apparently converging toward
a value near ${\rm [Fe/H]} = -0.4$ (e.g., Kuznetsov et al.\
\cite{Kuznetsov}).  Figure~\ref{fig:cmdra} presents the comparison
between the measured radii and temperatures of CM Dra as determined by
Morales et al.\ (\cite{Morales:09}) against isochrones from the
Dartmouth series by Dotter et al.\ (\cite{Dotter:08}) for two
metallicities and two different ages. One may draw two conclusions
from this diagram: 1) matching the radii and temperatures requires
models with an unusually high metallicity around ${\rm [Fe/H]} =
+0.5$, which is at odds with expectations; 2) despite claims often
seen in the literature, age is not totally irrelevant when comparing
slowly-evolving low-mass stars such as these with models. In fact, the
figure shows that changing the isochrone age from 4 to 10 Gyr leads to
a significant change in the predicted radii compared to the
uncertainties.

Similar discrepancies are seen for YY Gem and CU Cnc. These two are
particularly important systems because there is some additional (if
somewhat circumstantial) information on the age and metallicity that
eliminates free parameters when comparing with models. Another of the
well-studied systems, GU Boo, has components that are also too large
and too cool compared to predictions, and in this case even models
with a metallicity as high as ${\rm [Fe/H]} = +0.5$ do not quite match
the measured radii.

It has long been realized that essentially all of these discrepant
low-mass binary systems have short orbital periods, typically less
than 3 days. The leading hypothesis to explain the radius inflation
and temperature suppression has therefore been that tidal forces in
these tight systems tend to synchronize the stellar spins with the
orbital motion, resulting in rapid rotation and associated magnetic
activity. Activity is known to inhibit convective transport of energy
in the stellar interiors, and this in turn leads to larger radii and
cooler temperatures. There is at least a first-order theoretical
understanding of these processes (e.g., Gough \& Taylor
\cite{Gough:66}; Mullan \& MacDonald \cite{Mullan:01}; Chabrier,
Gallardo \& Baraffe \cite{Chabrier:07}; MacDonald \& Mullan
\cite{MacDonald:12}), as well as empirical evidence that the use of a
smaller mixing length parameter $\alpha_{\rm ML}$ in the models (to
emulate reduced convective efficiency) does indeed improve the fit to
the observations. Spot coverage associated with stellar activity
reduces the radiating surface area, and this can also contribute to
the discrepancies. Alternate explanations, such as errors in the
opacities or metallicity effects, seem less likely.

If the short orbital periods are at the root of the problem, then a
logical expectation would be that systems with longer periods would
agree with the models much better.  Tidal forces should be weaker, the
stars would presumably not be synchronized and should rotate more
slowly, and activity would therefore be much lower.  Unfortunately
long-period eclipsing binaries ($P > 10$ days) are rare and
exceedingly difficult to study (it can take several observing seasons,
good luck with the weather, and plenty of patience to complete a light
curve), and until recently there were no such systems with accurate
mass and radius determinations. In 2011 the MEarth and \emph{Kepler}
transit surveys reported the discovery of two eclipsing binaries with
low-mass components that coincidentally have the same orbital period
(about 41 days) to within 0.1\%. Contrary to expectations, the
low-mass stars in the MEarth system LSPM\,J1112+7626
(0.39\,$M_{\odot}$ and 0.27\,$M_{\odot}$; Irwin et al.\
\cite{Irwin:11}) both have inflated radii and are cooler than
predicted by theory. The secondary in Kepler-16 (0.20\,$M_{\odot}$;
Doyle et al.\ \cite{Doyle:11}; Winn et al.\ \cite{Winn:11}) is also
inflated. Its temperature has not been determined, and the primary is
a more massive K star. The activity level of Kepler-16\,B is unknown,
and there is perhaps some evidence of activity in the secondary of
LSPM\,J1112+7626, even at this long a period.

The situation regarding the nature of the discrepancies with models is
thus not as clear-cut as expected, and to make matters more
interesting, there is at least one well studied triple system from
\emph{Kepler} with two M-dwarf components (KOI-126; Carter et al.\
\cite{Carter:11}) in which the stellar radii seem to agree perfectly
with the Dartmouth models (see Feiden et al.\ \cite{Feiden:11}).

\section{Putting it all in together}
\label{sec:together}

The vast numbers of light curves produced by exoplanet transit surveys
and the increased interest in these disagreements between theory and
observation have led to a surge in low-mass eclipsing binary
discoveries in the last few years. It has become routine for each new
discovery paper to present a (typically well-populated) mass-radius
diagram highlighting the new system, and showing one or another set of
model isochrones for comparison. The objects displayed are usually
drawn from many different sources, sometimes including interferometric
measurements for single stars or results from eclipsing single-lined
spectroscopic binaries.  Authors have used such diagrams to draw
general conclusions about the magnitude of the radius anomalies, and
some studies have even proposed patterns in the discrepancies
depending on mass or period.

In a review on low-mass stars a few years ago, Ribas (\cite{Ribas:06})
provided an interesting illustration of the confusion that can result
from including M-dwarf systems indiscriminately in such mass-radius
diagrams. For a first version of the diagram he compiled and included
all low-mass systems known at the time (34 stars, including
single-lined binaries and stars with measured angular diameters), and
noted that on average the data appeared discrepant with the models
above the convective boundary ($\sim$0.35\,$M_{\odot}$), but seemed to
agree better with theory below that limit. He then restricted the
sample to only the double-lined eclipsing systems with mass and radius
errors under 3\%, and the deviations were seen much more clearly, all
the way to the lowest-mass system (CM Dra) below the convective
boundary.

\begin{figure}
\includegraphics[angle=0,width=\linewidth]{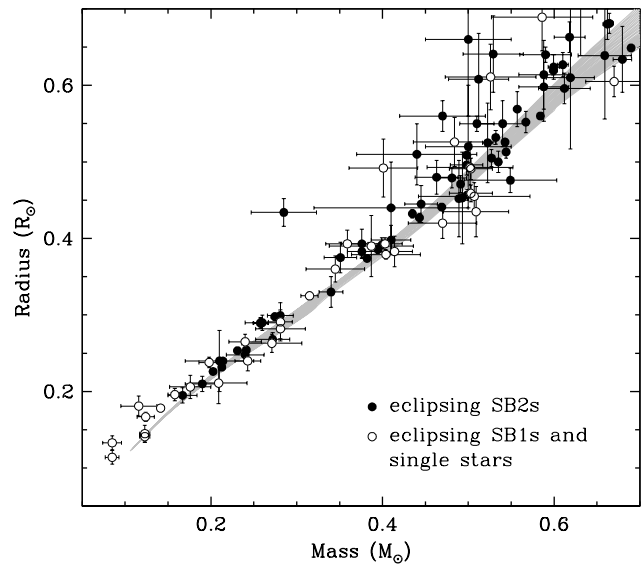}
\caption{Mass-radius diagram for low-mass stars, including all
measurements for double-lined eclipsing binaries (SB2s, filled
symbols) as well as determinations for single-lined eclipsing systems
(SB1s) and single stars (open symbols). Solar-metallicity Dartmouth
isochrones are shown for comparison, for ages ranging from 1 to 13 Gyr
(grey band).}
\label{fig:MRall}
\end{figure}
\begin{figure}
\includegraphics[angle=0,width=\linewidth]{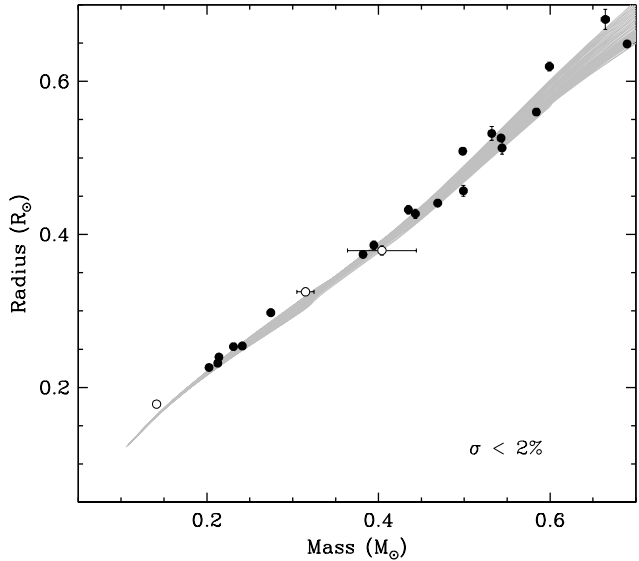}
\caption{Same as Fig.~\ref{fig:MRall}, limited to stars with mass and
radius errors under 2\% for double-lined eclipsing binaries, and
radius errors below that limit for single-lined systems and single
stars. No restriction was placed on the mass errors for the later two
classes of objects, as the masses are often adopted from a
mass-luminosity relation or rely on other assumptions.}
\label{fig:MR2percent}
\end{figure}

Many more determinations have become available since.
Figure~\ref{fig:MRall} shows an update of the mass-radius diagram that
includes all double-lined as well as single-lined eclipsing systems of
which we are aware with at least one component measured to be under
0.7\,$M_{\odot}$. In also includes single stars in this range with
radii determined interferometrically and masses inferred from a
mass-luminosity relation. There are a total of 108 stars displayed on
the graph.  For the vast majority the chemical composition is unknown,
so we have chosen to compare the observations with Dartmouth models
for solar metallicity. We also do not know the ages of most of these
stars, and as indicated earlier age can affect the radius in a
non-negligible way even for the slowly-evolving lowest-mass stars.
Some authors have compared the observations against a very young
isochrone (e.g., 300 Myr), but this is not necessarily a typical age
for a field star and will tend to exaggerate the radius
discrepancies. Therefore, we have chosen to display isochrones for all
ages from 1 to 13 Gyr. The model is thus represented by a band instead
of a single line.

The scatter in Figure~\ref{fig:MRall} is so large that very little can
be concluded from this diagram, and some may even be tempted to say
that there is nothing wrong with the models. This is of course not
true, as we have pointed out before. Restricting the sample to only
the stars with formal errors under 5\% (56 objects) doesn't change the
picture significantly. The radius discrepancies are still obscured by
the scatter and the uncertainties in the models. In
Figure~\ref{fig:MR2percent} we have been even more selective, setting
the error limit to 2\% (24 stars).  Given the age uncertainties, one
is still hard-pressed to draw any meaningful conclusions.

\begin{figure}
\includegraphics[angle=0,width=\linewidth]{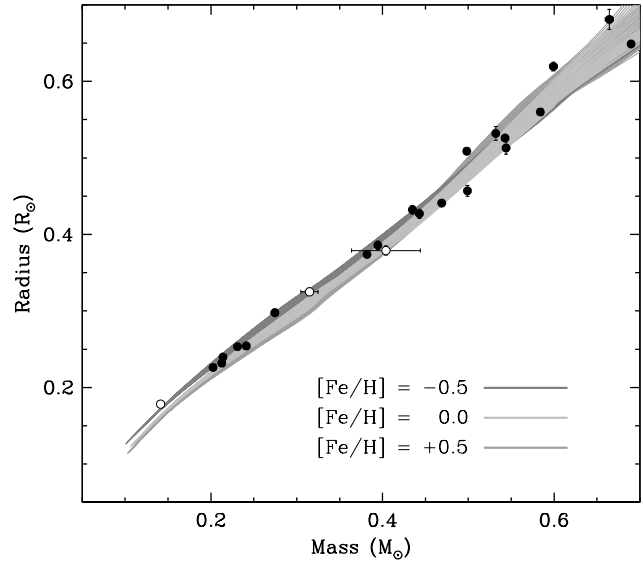}
\caption{Same as Fig.~\ref{fig:MR2percent}, but now showing model
isochrones for metallicities of ${\rm [Fe/H]} = -0.5$ and ${\rm
[Fe/H]} = +0.5$, in addition to solar, and all ages from 1 to 13 Gyr.}
\label{fig:MRmet}
\end{figure}

Stars are of course under no obligation to all have solar metallicity.
So as an exercise, if we now display models (all ages) for ${\rm
[Fe/H]} = -0.5$ and ${\rm [Fe/H]} = +0.5$, in addition to solar
composition, any systematic differences between the isochrones and the
observations appear even less obvious (Figure~\ref{fig:MRmet}).

\section{The scatter in the mass-radius diagram}
\label{sec:scatter}

In the last few years it has become increasingly clear (at least to
the author) that successive updates of the mass-radius diagram
featuring more and more low-mass stars have not necessarily led to a
deeper understanding of the problem. The scatter in the diagram has
become quite large, and the evidence for disagreements between theory
and observation, which is unmistakable when focusing on the best
studied \emph{individual} systems, is getting blurred when looking at
the larger sample of all available determinations.  This loss of
clarity from the increased dispersion is due to at least three causes:
1) published formal errors do not necessarily reflect the total
uncertainty; in fact, systematic errors tend to dominate in this mass
regime, as discussed further below; 2) in the great majority of cases
the age and metallicity are unknown; quantifying the radius
discrepancy by comparison with an arbitrarily chosen model can be
misleading; 3) the degree of radius inflation may not be the same for
all stars (recall KOI-126, which shows no such anomaly); it remains to
be confirmed whether this is a function of the strength of the
activity (which is not always a known property), stellar mass, or some
other parameter.

In a recent review on accurate stellar masses and radii, Torres et
al.\ (\cite{Torres:10}) examined the credentials of all known
double-lined eclipsing binaries studied up to that time, and presented
a short list of only four M-dwarf systems with well-measured
properties (relative errors of 3\% or less) that satisfied their
strict selection criteria. These four systems are CM Dra, YY Gem, CU
Cnc, and GU Boo.  Many authors have adopted similar cutoffs for the
errors when selecting new (or old) systems for the mass-radius
diagram, but have tended to overlook other selection criteria that are
perhaps more difficult to apply and require personal judgement, but
are just as important.  We suspect the masses and radii for many of
these often used systems may be biased, and can potentially lead to
more confusion. To state the obvious, precision is not the same as
accuracy.

In addition to setting an upper limit on the errors, it is important
to verify that the quality and quantity of the data used in the mass
and radius determinations are adequate to support the claimed
uncertainties (see Sect.~3 of Torres et al.\ \cite{Torres:10}, and
also Andersen \cite{Andersen:91}).  Incomplete light curves, such as
ones with no out-of-eclipse observations or that do not fully cover
both eclipses, can easily lead to systematic errors in the radii.
Similarly, radial-velocity curves based on only a handful of
observations, or with less-than-optimal phase coverage (e.g., missing
one or both quadratures) are also questionable, regardless of the
internal precision of the velocities.  The data analysis techniques
are important as well. Light curves often present strong degeneracies
among several of the fitted parameters, and it is all too easy to be
misled by the quick convergence of the solutions. In particular,
determining the individual radii (or their ratio) accurately in
partially eclipsing systems with nearly equal components can be
difficult, and in many cases requires the application of external
constraints such as a spectroscopic luminosity ratio to remove the
degeneracy (see Andersen \cite{Andersen:91}).

The most reliable analyses are those that pay close attention to all
of these issues, and document consistency checks or other efforts to
control or at least assess the impact of systematic errors. Although
many new mass and radius determinations for low-mass binary systems
have appeared in the literature since the compilation of Torres et
al.\ (\cite{Torres:10}), and are commonly shown in recent mass-radius
diagrams, we have a feeling that only a small number meet all of the
stringent criteria outlined above. Deciding which studies to trust
requires a critical review of the published analyses, a task that is
beyond the scope of this paper.

For the above reasons we believe we may have reached the point at
which it is more fruitful to compare well-measured individual systems
against models, rather than to include a larger number of more
questionable determinations and attempt to draw general quantitative
conclusions.

\section{Understanding the anomalies}
\label{sec:understanding}

A number of investigations, both observational and theoretical, have
examined possible correlations between the radius and temperature
anomalies and the strength of the stellar activity. On the
observational side, L\'opez-Morales (\cite{Lopez-Morales:07}) used the
X-ray luminosity as an activity indicator, and showed that for close
binary stars with low-mass components the degree of radius inflation
seems to depend on $L_{\rm X}/L_{\rm bol}$, while for single stars no
clear dependence was seen.  As others have pointed out, however, the
range of activity levels for the single stars in that study was rather
small, so the conclusion should be considered tentative pending
confirmation with a larger sample. Another study by Morales et al.\
(\cite{Morales:08}) focused on single stars of spectral type late K
and M, and used the strength of the H$\alpha$ emission to distinguish
active from inactive objects. They found that single active low-mass
stars show similar radius and temperature anomalies as the stars in
well-studied eclipsing binaries, implying that the problem is
independent of whether the object is in a binary or not.  More recently
Stassun et al.\ (\cite{Stassun:12}) also made use of H$\alpha$
measurements for single stars, as well as X-ray activity measurements
in binaries, and derived empirical relations between the radius
inflation and temperature suppression as a function of activity.

On the theoretical side, Mullan \& MacDonald (\cite{Mullan:01})
compared active and inactive single stars as measured by their X-ray
luminosity, and were able to explain their systematically different
global properties (temperatures and radii) both qualitatively and
quantitatively using custom models incorporating magnetic fields. They
parametrized them in terms of a magnetic inhibition parameter
$\delta$ (the ratio of the magnetic to total energy density) to
describe the reduction in the efficiency of convection in the stellar
interiors.  They recently applied the same scheme to successfully
model the components of CM Dra (MacDonald \& Mullan
\cite{MacDonald:12}).  Earlier studies such as that by D'Antona,
Ventura \& Mazzitelli (\cite{DAntona:00}) had also shown how magnetic
fields change the global structure of stars and how models that
incorporate these effects can be made to match the observations. More
recently Chabrier et al.\ (\cite{Chabrier:07}) were also able to
explain the radius and temperature anomalies in low-mass stars with a
different theoretical approach, and suggested that two different
manifestations of stellar activity can change the sizes and
temperatures of low-mass stars: magnetic inhibition, and spot
coverage.  They modeled both and showed that with a reduced mixing
length parameter $\alpha_{\rm ML}$ and a suitable spot-filling factor
$\beta$ (fraction of the star covered by spots) it is possible to
match the inflated radii and suppressed temperatures of these objects.

According to the above theoretical studies cool spots have a real and
detectable effect on the global stellar properties because they reduce
the effective radiating area, puffing up the star and reducing its
temperature. But they can also have a deleterious effect on the
measurements, specifically on the radius determinations. Spots
typically cause modulations and/or features in the light curves,
affecting the shape of the eclipses.  Numerical simulations carried
out by Morales et al.\ (\cite{Morales:10}) showed that non-negligible
biases in the radii can result, depending on the spot distribution.
They found that polar spots (the kind often expected in rapidly
rotating stars, and actually seen in Doppler tomography) have the
largest effect, causing the radius to be overestimated by up to
3--6\%. Note that this bias goes in the same direction as the observed
discrepancies between models and observations, and would thus tend to
alleviate the differences if we were able to avoid it. The effect on
the mass determinations, on the other hand, was found to be less
serious (0.5--1\%). For spot distributions less concentrated to the
poles the effects on the radius are smaller and more random in nature.

These systematic errors are especially worrisome when considering that
spots tend to change or to come and go on active stars. A recent
illustration of this was given by Windmiller, Orosz \& Etzel
(\cite{Windmiller:10}), who re-observed the well-studied system GU Boo
(L\'opez-Morales \& Ribas \cite{Lopez-Morales:05}). They reanalyzed
the original light curves, as well as two new ones they obtained two
years later, and found differences in the measured radii at the level
of 2\%. They attributed most of this to the spots (with perhaps some
contribution from the modeling technique), which had changed visibly
in the intervening years.

The lesson from the empirical and numerical evidence described above
is that spots can cause systematic measurement errors in the radii at
the level of several percent, if the stars are sufficiently active.
This certainly complicates the picture, and suggests the need to
investigate these effects carefully for each new system.  At the very
least, it may argue for being more conservative in stating one's
measurement uncertainties for low-mass stars.

\section{Final remarks}
\label{sec:remarks}

The last few years have seen excellent progress in measuring
fundamental properties of low-mass stars, and in understanding the
underlying causes of the radius and temperature discrepancies with
standard (non-magnetic) models.  Recent models that incorporate the
effects of magnetic fields are able to match the measured properties
of low-mass stars, at the expense of adding one or two more free
parameters. It remains to be seen if these or other equivalent
parameters can be put in terms of some easily measurable quantity,
which would then allow more stringent tests of theory.

With transiting exoplanet and other photometric surveys in full swing,
prospects are good that the empirical evidence will continue to
build. The most useful studies will be those using complete and
high-quality data analyzed with appropriate techniques, and that pay
careful attention to systematic errors, especially those related to
spots. In additional to measuring masses and radii in eclipsing
binaries, observers should endeavor to determine also the effective
temperatures and metallicity whenever possible, and to estimate the
activity levels, which clearly play an important role.

\acknowledgements The author wishes to thank the organizers of the
meeting for the invitation to speak. This work was partially supported
by grant AST-10-07992 from the US National Science Foundation.


\end{document}